\documentclass{cmspaper}
\usepackage{epsfig}
\usepackage{amsmath}
\begin{document}
 
\renewcommand{\arraystretch}{1.5}

\begin{titlepage}

\flushright{\bf\Large IEKP-KA/2001-23}

   \cmsnote{2001/054}
   \date{15 November 2001}

  \boldmath
  \title{Searching for Higgs Bosons in Association with Top Quark Pairs in the $H^0 \rightarrow b\bar{b}$ Decay Mode}
  \unboldmath

  \begin{Authlist}
    V.~Drollinger and Th.~M\"uller
       \Instfoot{iekp}{IEKP, Karlsruhe University, Germany}
    D.~Denegri
       \Instfoot{cern}{CERN. Geneva, Switzerland and DAPNIA Saclay, France}
  \end{Authlist}


  \begin{abstract}

Search for the Higgs Boson is one of the prime goals of the LHC. Higgs bosons lighter than 130~$GeV/c^2$ decay mainly to a $b$-quark pair. While the detection of a directly produced Higgs boson in the $b\bar{b}$ channel is impossible because of the huge QCD background, the channel $t\bar{t} H^0 \rightarrow l^\pm \nu q\bar{q} b\bar{b} b\bar{b}$ is very promising in the Standard Model and the MSSM.

We discuss an event reconstruction and selection method based on likelihood functions. The CMS detector response is performed with parametrisations obtained from detailed simulations. Various physics and detector performance scenarios are investigated and the results are presented. It turns out that excellent $b$-tagging performance and good mass resolution are essential for this channel.

  \end{abstract} 

\newpage
\ 

  
\end{titlepage}

\section{Introduction}

\begin{figure}[ht]
\begin{center}
 \includegraphics[width=0.40\textwidth,angle=+0]{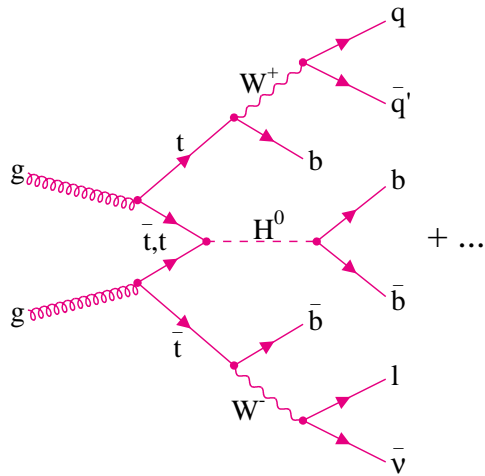}
 \caption{\sl One example of a $t\bar{t} H^0 \rightarrow l^\pm \nu q\bar{q} b\bar{b} b\bar{b}$ event at LO.\rm}
 \label{fig:fey_tth}
\end{center}
\end{figure}

The Higgs mechanism \cite{HiggsMech} is the generally accepted way to generate particle masses in the electroweak theory. If the Higgs boson is lighter than 130 $GeV/c^2$, it decays mainly to a $b\bar{b}$ pair \cite{HDECAY}. To observe the Higgs boson at the LHC, the $t\bar{t}H^0$ channel turns out to be the most promising channel among the Higgs production channels with $H^0 \rightarrow b\bar{b}$ decay \cite{THESIS}. In this study, we discuss the channel $t\bar{t} H^0 \rightarrow l^\pm \nu q\bar{q} b\bar{b} b\bar{b}$ (Figure~\ref{fig:fey_tth}), where the Higgs Boson decays to $b\bar{b}$, one top quark decays hadronically and the second one leptonically. The relevant signal and background cross sections at the LHC ($\sqrt{s_{pp}} = $ 14 $TeV$) and particle masses used in the simulation are listed in 
\vspace*{5mm}
\begin{table}[ht]
 \begin{center}
 \begin{tabular}{|lcr|lcr|}
 \hline
 \multicolumn{3}{|c|}{LO cross sections} & \multicolumn{3}{|c|}{masses}\\
 \hline
  $\sigma_{t\bar{t}H^0} \times BR_{H^0 \rightarrow b\bar{b}}$ & = 
 & 1.09 - 0.32 $pb$ &  $m_{H^0}$ & = & 100 - 130 $GeV/c^2$ \\
    $\sigma_{t\bar{t}Z^0}$ & = & 0.65 $pb$ & $m_{Z^0}$ & = & 91.187 $GeV/c^2$\\
 $\sigma_{t\bar{t}b\bar{b}}$ & = & 3.28 $pb$ &   $m_{b}$ & = & 4.62 $GeV/c^2$\\
       $\sigma_{t\bar{t}jj}$ & = & 507  $pb$ &   $m_{t}$ & = &  175 $GeV/c^2$\\
 \hline
 \end{tabular}
 \end{center}
  \caption{\sl CompHEP \cite{CompHEP} cross sections for signal and background relevant for the $t\bar{t} H^0 \rightarrow l^\pm \nu q\bar{q} b\bar{b} b\bar{b}$ channel, calculated with parton density function CTEQ4l \cite{PDFLIB}. The branching ratio of the semileptonic decay mode (one $W^\pm$ decays to quarks the other $W^\pm$ decays leptonically, where only decays to electrons or muons are taken into account) is 29\%  (not included in the cross sections of this table) and $m_{W^\pm} =$ 80.3427 $GeV/c^2$.\rm}
\label{tab:crossections}
\end{table}
Table~\ref{tab:crossections}. The hard processes are generated with CompHEP and then interfaced to PYTHIA, where the fragmentation and hadronisation are performed \cite{CompHEP}-\cite{PYTHIA}. After the final state including the underlying event has been obtained, the CMS detector response is simulated, with track and jet reconstruction with parametrisations FATSIM \cite{FATSIM} and CMSJET \cite{CMSJET}, obtaining in this way tracks, jets, leptons (the electron or muon reconstruction efficiency is assumed to be 90\%; taus are not considered here) and missing transverse energy. These parametrisations have been obtained from detailed simulations based on GEANT.
\vspace*{0mm}
\vspace*{-2mm}


\section{Reconstruction}

From Figure~\ref{fig:fey_tth} we expect to find events with one isolated lepton, missing transverse energy $E_T^m$ and six jets (four $b$-jets and two non-$b$-jets), but initial and final state radiation are sources of additional jets. So the number of jets per event is typically higher than six. On the other hand, not all six quarks of the hard process can be always recognised as individual jets in the detector, in which case it is impossible to reconstruct the event correctly - even if there are six or more jets.

For the reconstruction of resonances it is necessary to assign the $n$ jets of an event to the corresponding quarks of the hard process. In general, and ignoring information on $b$-jets, the number of possible combinations $N$ is given in Table~\ref{tab:combinations} as a function of the number of jets per event. We obtain $N$ for the case, when the masses of the Higgs boson, both top quarks and the hadronically decaying W boson are reconstructed. The nominal mass of the leptonically decaying W boson, together with $E_T^m$ and the lepton four momentum, is used to calculate two solutions of the longitudinal momentum of the neutrino $p_Z(\nu)$ which is needed for the mass reconstruction of the leptonically decaying top.

Good mass resolution and the identification of $b$-jets is essential to reduce the number of wrong combinations in the event reconstruction. A good mass resolution can be obtained when the energy and direction of each reconstructed jet agree as closely as possible with the quantities of the corresponding parent quark. This can be achieved with jet corrections as described in \cite{JETCOR} and \cite{JETRAD}. For $b$-tagging we use $b$-probabilities (\ref{B-PROB}) and (\ref{L-PROB}) (see appendix) which depend on impact parameters of tracks and leptons inside the jets. They are determined using $t\bar{t}$ six jet events, as described in \cite{THESIS}. The identification of $b$-jets is even more important for efficient background suppression.
\vspace*{3mm}
\begin{table}[ht]
 \begin{center}
 \begin{tabular}{|ccccccccc|}
 \hline
 \multicolumn{9}{|c|}{\rule[-3mm]{0mm}{8mm} $N = \binom{n}{6} \times 6! \times \frac{1}{2} \times \frac{1}{2} \times 2\ = \binom{n}{6} \times 360$}\\
 \hline
 $n$ & = &   6 &    7 &     8 &     9 &    10 &     11 &     12 \\
 $N$ & = & 360 & 2520 & 10080 & 30240 & 75600 & 166320 & 332640 \\
 \hline
 \end{tabular}
 \end{center}
  \caption{\sl Number of jets per event $n$ and the corresponding number of possible combinations $N$. If there are more than a dozen jets, only the twelve with highest $E_T$ are considered.\rm}
\label{tab:combinations}
\end{table}

\begin{figure}[ht]
\begin{center}
 \includegraphics[width=0.70\textwidth,angle=+0]{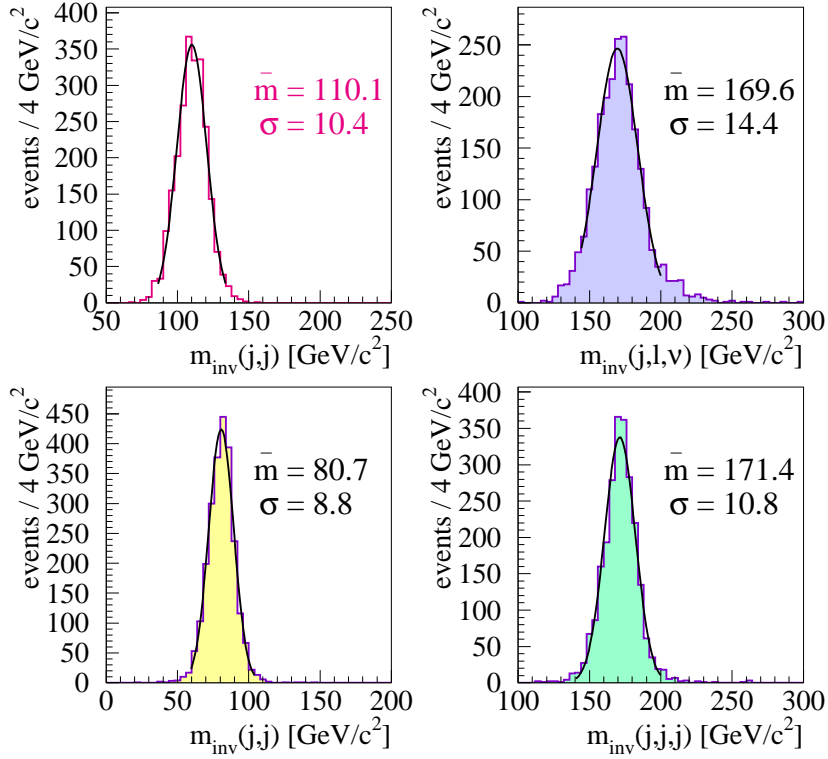}
 \caption{\sl Invariant resonance masses of the $t\bar{t} H^0 \rightarrow l^\pm \nu q\bar{q} b\bar{b} b\bar{b}$ signal: Higgs boson, leptonic top, hadronic top and hadronic $W^\pm$. The leptonic $W^\pm$ is not reconstructed but its nominal mass is used to calculate $p_Z(\nu)$. The generated masses are: $m_{H^0} =$ 115 $GeV/c^2$, $m_{t} =$ 175 $GeV/c^2$ and $m_{W^\pm} =$ 80.3427 $GeV/c^2$.\rm}
 \label{fig:resonances}
\end{center}
\end{figure}

Figure~\ref{fig:resonances} shows the invariant mass distributions of the reconstructed resonances of $t\bar{t} H^0 \rightarrow l^\pm \nu q\bar{q} b\bar{b} b\bar{b}$ events in the case of an ideal reconstruction: after the ``preselection'' and the calculation of $p_Z(\nu)$ (see later on) each quark of the hard process is matched with exactly one jet, the closest one in $R = \sqrt{\phi^2 + \eta^2}$ if $\Delta R(q,j) <$ 0.3 and if the jet energy is closer than $\pm$ 30 \% to the parent quark energy. The mean values and widths of the top and W mass distributions are used to define likelihood functions used in the selection procedure described in the following.

{\bf\boldmath$\diamond$ Preselection}\\
Events are selected if there is an isolated lepton ($e^\pm$ or $\mu^\pm$ with $p_T >$ 10 $GeV/c$ within the tracker acceptance; no other track with $p_T >$ 1 $GeV/c$ in a cone of 0.2 around the lepton) and at least six jets ($E_T >$ 20 $GeV$ , $|\eta| <$ 2.5).

{\bf\boldmath$\diamond$ Event Configuration}\\
In order to be able to reconstruct the Higgs mass, we have to find the correct event configuration among all possible combinations listed in Table~\ref{tab:combinations}. The best configuration is defined as the one which gives the highest value of an event likelihood function (\ref{EV-L-SHORT}) which takes into account $b$-tagging of four jets, anti-$b$-tagging of the two jets supposed to come from the hadronic $W^\pm$, mass reconstruction of $W^\pm$ and the two top quarks, and sorting of the $b$-jet energies.
\begin{align}
 {\it\bf L\_EVNT} & = \prod_{i=1,4}P_b(b_i) 
               \times \prod_{i=1,2}[1 - P_b(q_i)]
               \times \prod_{i=W^\pm ,t,\bar{t}}e^{-0.5 \times [\frac{m_i - \bar{m_i}}{\sigma_i}]^2}
               \times f[E_b(t,\bar{t})-E_b(H^0)]
\label{EV-L-SHORT}
\end{align}
The detailed version of this event likelihood function can be found in the appendix.

{\bf\boldmath$\diamond$ Jet Combinations}\\
Events with more than six jets can contain gluon jets from final state radiation, which are not yet used in the analysis. The combination of these jets with the correct quark jets can improve the event reconstruction further. The additional jets are combined with the decay products of both top quarks if they are closer than $\Delta R(j,j) <$ 1.7, if the corresponding mass is closer to the expected value of Figure~\ref{fig:resonances}. If there are still jets left, they are considered as Higgs decay products and are combined with the closest of the corresponding two $b$-jets, if $\Delta R(j,j) <$ 0.4. 

{\bf\boldmath$\diamond$ Event Selection}\\
Three likelihood functions: for resonances (\ref{TTH-L-RESO}) ($L\_RESO >$ 0.05), $b$-tagging (\ref{TTH-L-BTAG}) ($L\_BTAG >$ 0.50), and kinematics (\ref{TTH-L-KINE}) ($L\_KINE >$ 0.2) are used to reduce the fraction of background events. Finally, the events are counted in a mass window around the expected Higgs mass peak ($m_{inv}(j,j)$ in $\bar{m}\ \pm$ 1.9 $\sigma$ ; $\bar{m}$ and $\sigma$ are obtained from mass distributions as shown in Figure~\ref{fig:resonances} with various generated Higgs masses). The likelihood cuts have been optimised assuming a Higgs mass of 120 $GeV/c^2$.

\begin{figure}[ht]
\begin{center}
 \includegraphics[width=0.55\textwidth,angle=+0]{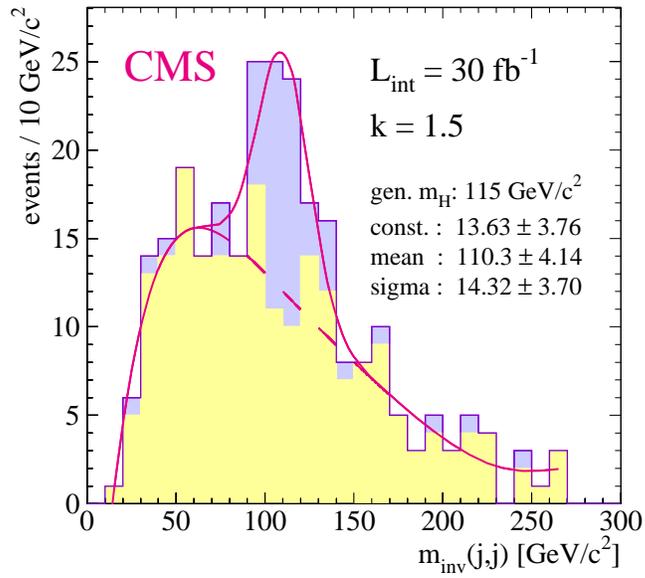}
 \caption{\sl Simulated invariant mass distribution of signal (dark shaded, $m_{H^0} =$ 115 $GeV/c^2$) plus background for $L_{int} = $ 30 $fb^{-1}$. The dashed curve is obtained from the fit of the background without signal, the solid line describes the fit of signal plus background.\rm}
 \label{fig:tth_sum115}
\end{center}
\end{figure}

The overall efficiency for a triggered event to be finally selected is 1.3\% for $t\bar{t}H^0$ ($m_{H^0} =$ 115 $GeV/c^2$), 0.2\% for $t\bar{t}Z^0$, 0.4\% for $t\bar{t}b\bar{b}$ and 0.003\% for $t\bar{t}jj$ events. This shows that the reducible background is reduced very effectively. In addition, there is little combinatorial background left (an example is shown in Figure~\ref{fig:tth_sum115}) with this reconstruction method.

\vspace*{5mm}


\section{SM Results}

\begin{figure}[ht]
\begin{center}
 \includegraphics[width=0.66\textwidth,angle=+0]{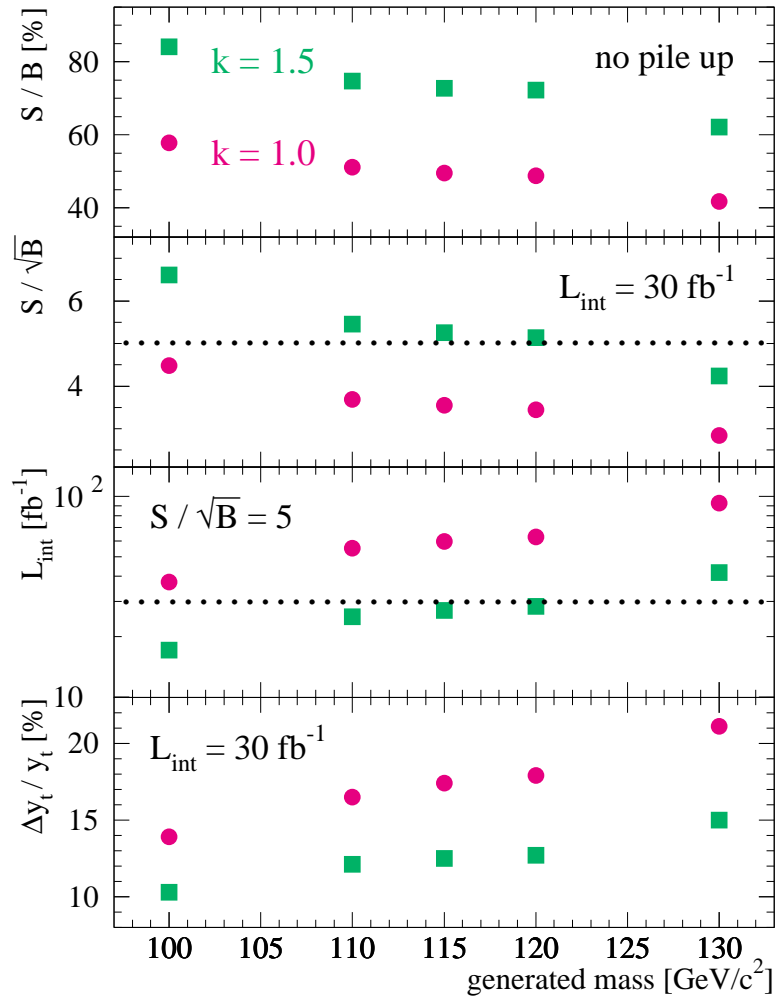}
 \caption{\sl $S / B$, $S / \sqrt{B}$, $L_{int}$ (required for $S / \sqrt{B} =$ 5) and $\Delta y_t / y_t$ versus generated Higgs mass in the SM. Two k-factor scenarios (k$_{t\bar{t} H^0 ,\ t\bar{t} Z^0} =$ k and k$_{t\bar{t} q\bar{q}} =$ 1.9) are shown: k = 1.0 (dots) and k = 1.5 (boxes).\rm}
 \label{fig:tth_stat4}
\end{center}
\end{figure}

After the whole reconstruction and event selection procedure, it turns out that the irreducible background (with four real $b$-jets) is dominant. Even the $t\bar{t}jj$ background, where only two $b$-jets from the top decays are generated in the hard process, is dominated by events with four real $b$-jets. This is possible after the fragmentation of PYTHIA: e.g. $gg \rightarrow t\bar{t} gg \rightarrow l^\pm \nu q\bar{q} b\bar{b} g b\bar{b}$ with one $b\bar{b}$ pair coming from $g \rightarrow b\bar{b}$ (gluon splitting). In this case the final state consists of nine partons or leptons which is one more than expected at LO and is therefore considered as HO (in this case NLO) process. Together with the number of $t\bar{t}b\bar{b}$ events (considered as LO) we obtain an intrinsic k-factor k$_{t\bar{t} q\bar{q}} =$ 1.9 for all $t\bar{t} q\bar{q}$ events. For the $t\bar{t}H^0$ signal and the $t\bar{t}Z^0$ background we assume two scenarios for k$_{t\bar{t} H^0 ,\ t\bar{t} Z^0} =$ k: LO (no k-factor) k = 1.0 and a more optimistic case k = 1.5. In the meanwhile a NLO calculation of the $t\bar{t}H^0$ cross section has been performed \cite{K-FAK}, where k $\approx$ 1.2 at a central scale $\mu = (2m_t + m_{H^0})/2$. For these two k-factor scenarios, the signal to background ratio $S / B$, the significance $S / \sqrt{B}$ for $L_{int} = $ 30 $fb^{-1}$, the integrated luminosity $L_{int}$ required for a significance of five or more and the precision on $y_t$ for $L_{int} = $ 30 $fb^{-1}$ are shown in Figure~\ref{fig:tth_stat4} as a function of the generated Higgs mass.
\vspace*{2mm}

$S / B$ is of the order of 50\% or higher, the significance is relatively high already for $L_{int} = $ 30 $fb^{-1}$, and the significance is above five for a low integrated luminosity. An integrated luminosity $L_{int} = $ 100 $fb^{-1}$ would be enough to explore all points considered in Figure~\ref{fig:tth_stat4}. Apart from these results, the Higgs mass can be determined from the Gaussian fit of the final mass distribution (see Figure~\ref{fig:tth_sum115}) with a precision of better than 4\%. Finally, the total event rate determines the top Higgs Yukawa coupling $y_t$ with a precision of around 15\%, if we assume a known branching fraction of the decay $H^0 \rightarrow b\bar{b}$.

\section{MSSM Results}

To give an idea about the discovery potential of the corresponding channel $t\bar{t} h^0 \rightarrow l^\pm \nu q\bar{q} b\bar{b} b\bar{b}$ in the MSSM, we extrapolate the SM results (by rescaling the production cross section times branching ratio, obtained with SPYTHIA \cite{SPYTHIA}) and discuss the parameter space coverage of one benchmark scenario called "maximum $m_h$" scenario \cite{SUSYbenchmark} which turns out to be the most difficult scenario. The reason is the rapidly falling cross section and branching ratio with increasing Higgs mass, which limits the discovery potential of this channel in the SM as well.
\vspace*{3mm}
\begin{figure}[ht]
\begin{center}
 \includegraphics[width=0.49\textwidth,angle=+0]{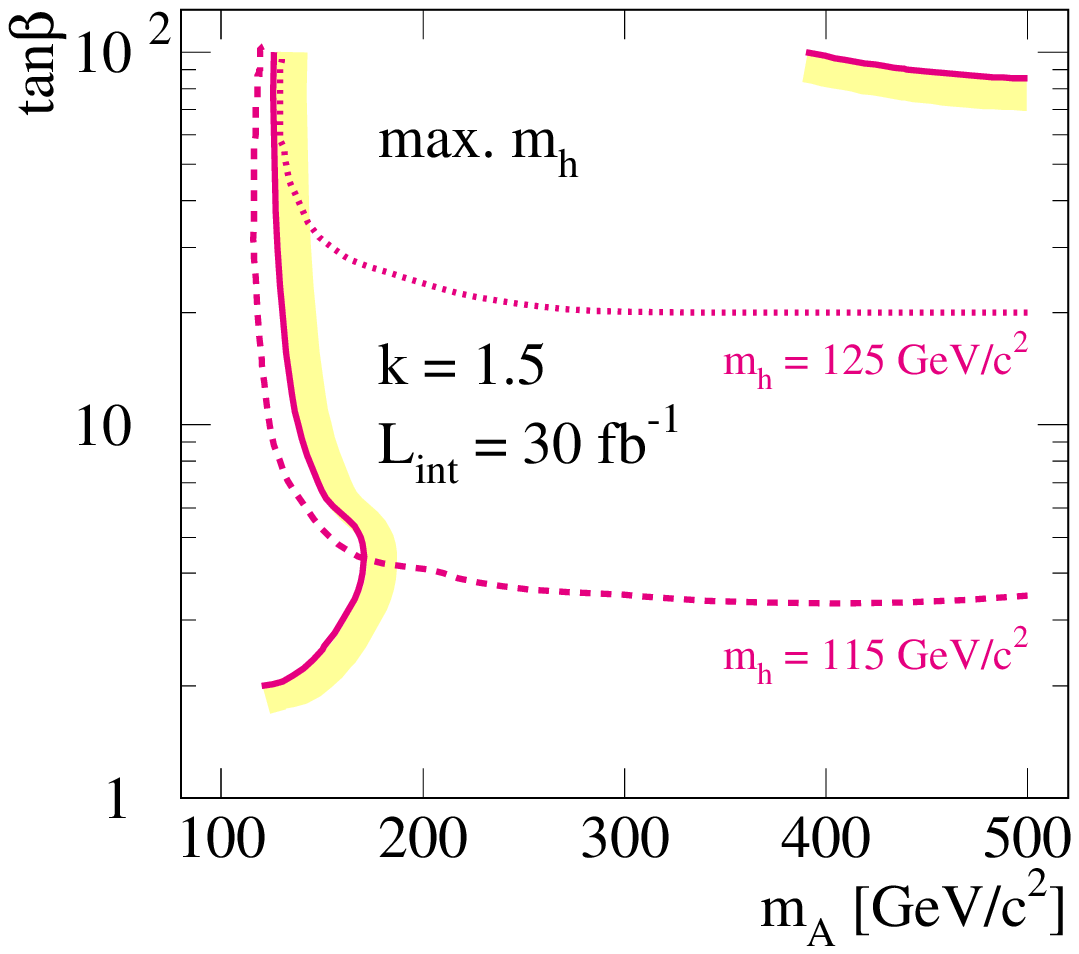}
 \includegraphics[width=0.49\textwidth,angle=+0]{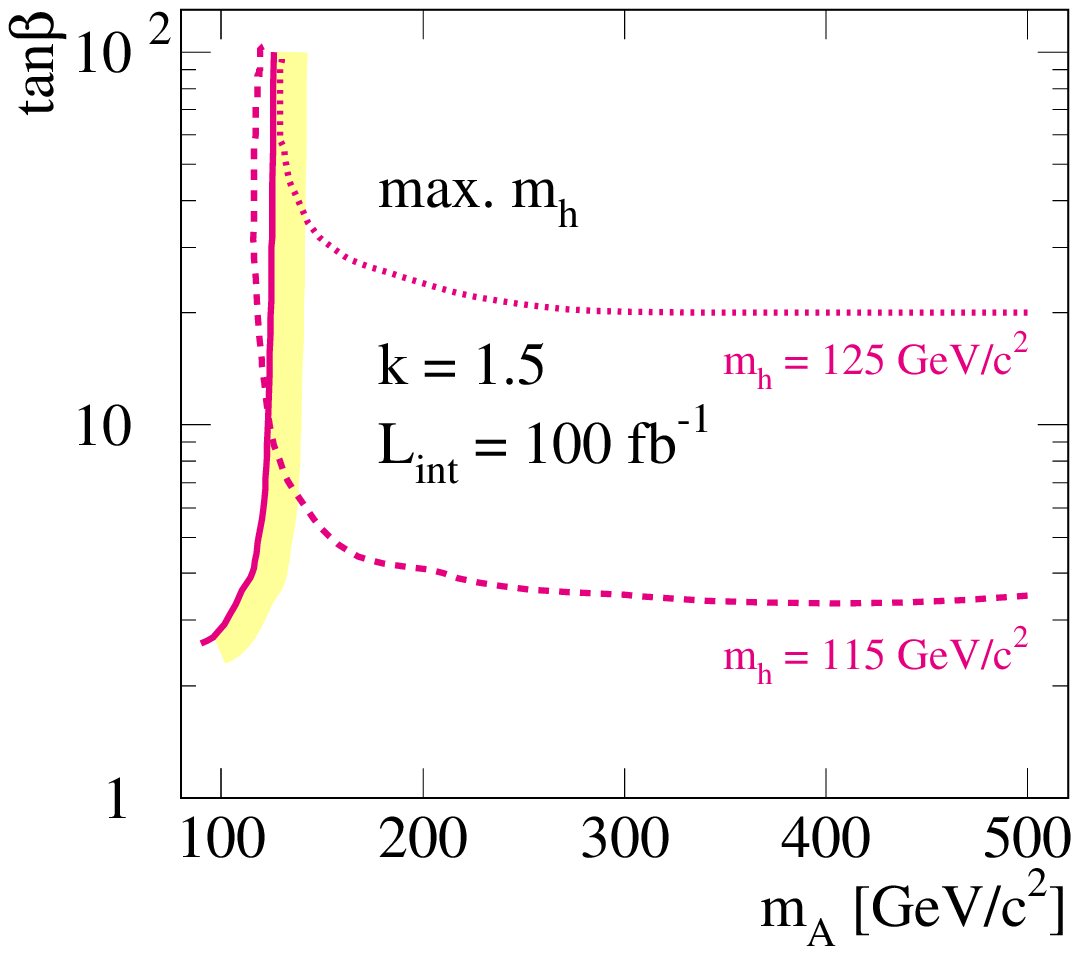}
 \caption{\sl Discovery contours in the MSSM ("maximum $m_h$" scenario) parameter space for $L_{int} = $ 30 $fb^{-1}$ (left) and for $L_{int} = $ 100 $fb^{-1}$ (right). $S / \sqrt{B} \ge$ 5 to the shaded side of the solid line. The dotted and dashed lines are the isomass curves for $m_{h^0} =$ 125 $GeV/c^2$ and $m_{h^0} =$ 115 $GeV/c^2$, respectively.\rm}
 \label{fig:tth_susy}
\end{center}
\end{figure}

Figure~\ref{fig:tth_susy} shows the parameter space coverage in the $m_A$-$\tan\beta$ plane for two integrated luminosities. In both cases there is an inaccessible region at low $m_A$, whereas the second difficult region at high $m_A$  and $\tan\beta$ disappears with increased integrated luminosity. In other scenarios the difficult regions are smaller, which means that for sufficient integrated luminosity most of the MSSM parameter space can be covered with this channel.


\section{Some CMS Performance Considerations}

We have obtained the previous results by considering jets with $|\eta| <$ 2.5 and $b$-tagging using both impact parameter measurements and the additional information on leptons ($e^\pm$ or $\mu^\pm$) inside the jets. For this particular scenario the result is shown again in Table~\ref{tab:performance} (second line). In the same table we compare the situation, when some of this information is not available. The first line is the result for the case when the information of leptons inside jets is missing. The $S/B$ is even somewhat higher, which means a higher purity, but the efficiency and the resulting significance are (not dramatically) lower. 
\vspace*{5mm}
\begin{table}[ht]
 \begin{center}
\begin{tabular}{|l|c|cccc|}
\hline
 $b$-tagging scenario & jet acceptance & $S$ & $B$ & $S/B$ & $S/\sqrt{B}$ \\
\hline
 without lepton information & $|\eta| <$ 2.5 & 26 & 31 & 84\% & 4.7 \\
 with lepton information    & $|\eta| <$ 2.5 & 38 & 52 & 73\% & 5.3 \\
 with lepton information    & $|\eta| <$ 2.0 & 30 & 41 & 75\% & 4.8 \\
 with lepton information    & $|\eta| <$ 1.5 & 20 & 27 & 73\% & 3.8 \\
\hline
\end{tabular}
 \end{center}
\caption{\sl Signal and background dependence on $b$-tagging scenario and jet acceptance, respectively. The numbers are given for $L_{int} = $ 30 $fb^{-1}$, k = 1.5 and $m_{H^0} =$ 115 $GeV/c^2$ in the SM.\rm}
\label{tab:performance}
\end{table}

In case of a reduced jet $\eta$ acceptance or tracker acceptance, respectively, signal and background are reduced in the same way. This gives practically constant $S/B$ and decreasing $S/\sqrt{B}$ for smaller acceptances. The result of the third line ($|\eta| <$ 2.0) is still good, but for an acceptance of $|\eta| <$ 1.5 (last line) the result is significantly worse. Because the signal to background ratio is stable, these effects can be compensated with higher integrated luminosity.


\section{Conclusions}

After a detailed study \cite{THESIS} we conclude that it is possible to reconstruct the $t\bar{t} H^0 \rightarrow l^\pm \nu q\bar{q} b\bar{b} b\bar{b}$ signal without significant combinatorial background, although effects of event pile up have still to be evaluated. There are two basic requirements: good jet reconstruction which guarantees a good mass resolution and excellent $b$-tagging performance which allows efficient and clean identification of $b$-jets. This helps to reduce the background substantially.

In the SM, a discovery is possible already after a short period of data taking at the LHC. The same is true in the MSSM, where most of the parameter space can be covered with the low integrated luminosity.

Beside the discovery of the Higgs boson, measurements of the Higgs mass and of the top Higgs Yukawa coupling are possible with considerable precision, which is important to understand the nature of the Higgs boson.

It is encouraging to see that in less favourable $b$-tagging conditions and with reduced acceptance the reconstruction of this channel does not break down altogether and the same results can be obtained by just increasing the integrated luminosity.


\appendix

\section{\boldmath $b$-Quark Distributions}

Transverse energy and pseudorapidity distributions for $b$-jets in $t\bar{t} H^0$ final states: Figure~\ref{fig:{kine_acc}}.

\vspace*{1mm}
\begin{figure}[ht]
\begin{center}
 \includegraphics[width=0.80\textwidth,angle=+0]{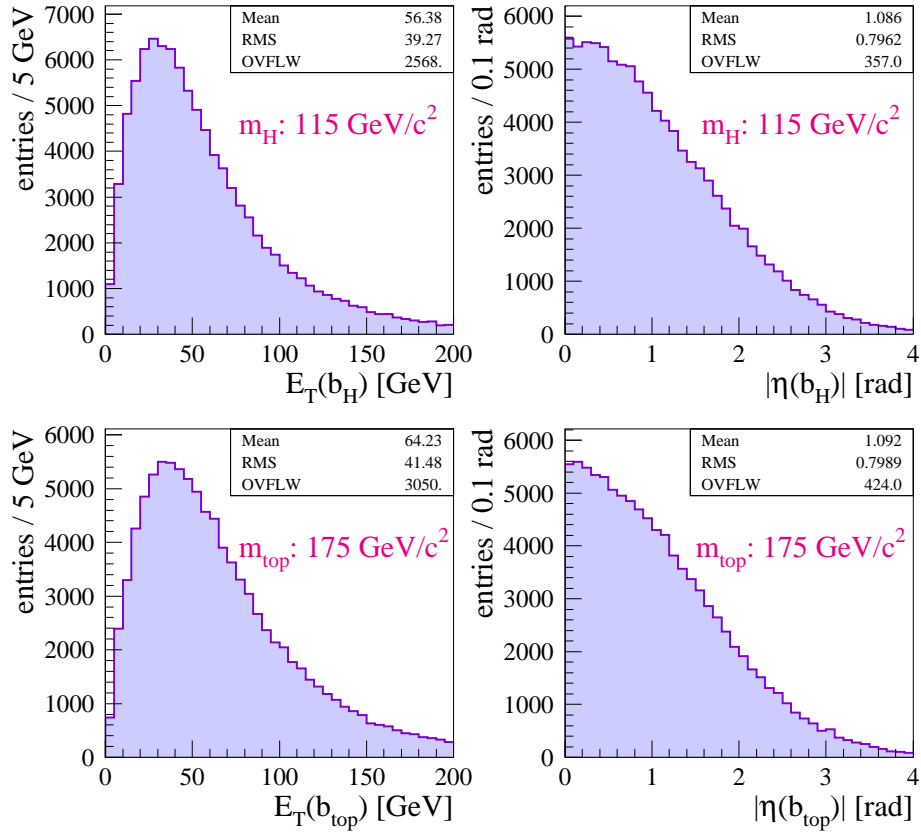}
 \caption{\sl $b$-quark $E_T$ (left) and $|\eta|$ (right) distributions obtained from the $t\bar{t} H^0 \rightarrow l^\pm \nu q\bar{q} b\bar{b} b\bar{b}$ signal without cuts: $b$-quarks from Higgs decay with $m_{H^0} =$ 115 $GeV/c^2$ are shown in the upper plots and $b$-quarks from top decays with $m_{t} =$ 175 $GeV/c^2$ are shown in the corresponding lower plots.\rm}
 \label{fig:{kine_acc}}
\end{center}
\end{figure}


\section{\boldmath $b$-Probability Functions}

The $b$-probability functions are used to define the likelihood functions (\ref{TTH-L-EVNT}) and (\ref{TTH-L-BTAG}). If there is a lepton reconstructed inside the jet, the $b$-probability is calculated from (\ref{L-PROB}), otherwise the function for jets without leptons (\ref{B-PROB}) is used.

\begin{align}
\begin{split}
 {\it\bf B\_PROB} & = \arctan[2.249 \sigma(ip) - 3.197 - 0.007709 E_T\\
                  & - \exp(0.7053 - 0.06249 E_T)] \times 0.2921 + 0.4877
\end{split}\label{B-PROB}\\
\intertext{}
\begin{split}
 {\it\bf L\_PROB} & = \arctan[1.510 \sigma(ip) - 1.394 - 0.008196 E_T\\
                  & - \exp(0.7624 - 0.08526 E_T)] \times 0.1026 + 0.8363
\end{split}\label{L-PROB}
\end{align}


\section{Likelihood Functions}

Likelihood functions which are used for the physics analysis of the $t\bar{t} H^0 \rightarrow l^\pm \nu q\bar{q} b\bar{b} b\bar{b}$ channel are defined in the following expressions. (\ref{TTH-L-EVNT}) is used to find the correct event configuration. The boldface variables represent the jets of an event. All possible combinations are checked: for instance, the jet with highest $E_T$ is treated as ${\bf BT_H}$ ($b$- jet of hadronic top decay), then it is treated as ${\bf B_1H}$ ($b$-jet "one" of the Higgs decay), then it is treated as ${\bf J_1W}$ ...
All other likelihood functions are defined for one (the final) event configuration.

\begin{align}
\begin{split}
 {\it\bf L\_EVNT} &      = b\text{-probability}[\sigma_{ip}({\bf B_1H}),E_T({\bf B_1H})]\\
                  & \times b\text{-probability}[\sigma_{ip}({\bf B_2H}),E_T({\bf B_2H})]\\
                  & \times b\text{-probability}[\sigma_{ip}({\bf BT_L}),E_T({\bf BT_L})]\\
                  & \times b\text{-probability}[\sigma_{ip}({\bf BT_H}),E_T({\bf BT_H})]\\
                  & \times (1 - b\text{-probability}[|\sigma_{ip}({\bf J_1W})|,E_T({\bf J_1W})])\\
                  & \times (1 - b\text{-probability}[|\sigma_{ip}({\bf J_2W})|,E_T({\bf J_2W})])\\
                  & \times \exp[ -0.5 \times \{\frac{m_{inv}({\bf BT_L},l,\nu)-169.6}{14.4}\}^2\ ]\\
                  & \times \exp[ -0.5 \times \{\frac{m_{inv}({\bf J_1W,J_2W})-80.7}{8.8}\}^2\ ]\\
                  & \times \exp[ -0.5 \times \{\frac{m_{inv}({\bf BT_H,J_1W,J_2W})-171.4}{10.8}\}^2\ ]\\
                  & \times \left[ \arctan\left(\frac{4[E({\bf BT_L}) + E({\bf BT_H}) - E({\bf B_1H}) - E({\bf B_2H})]}{E({\bf BT_L}) + E({\bf BT_H}) + E({\bf B_1H}) + E({\bf B_2H})}\right) \times \frac{1}{\pi} + \frac{1}{2} \right]
\end{split}\label{TTH-L-EVNT}\\
\intertext{}
\begin{split}
 {\it\bf L\_RESO} &      = \exp[ -0.5 \times \{\frac{m_{inv}(BT_L,l,\nu)-169.6}{14.4}\}^2\ ]\\
                  & \times \exp[ -0.5 \times \{\frac{m_{inv}(J_1W,J_2W)-80.7}{8.8}\}^2\ ]\\
                  & \times \exp[ -0.5 \times \{\frac{m_{inv}(BT_H,J_1W,J_2W)-171.4}{10.8}\}^2\ ]
\end{split}\label{TTH-L-RESO}\\
\intertext{}
\begin{split}
 {\it\bf L\_BTAG} &      = b\text{-probability}[\sigma_{ip}(B_1H),E_T(B_1H)]\\
                  & \times b\text{-probability}[\sigma_{ip}(B_2H),E_T(B_2H)]\\
                  & \times b\text{-probability}[\sigma_{ip}(BT_L),E_T(BT_L)]\\
                  & \times b\text{-probability}[\sigma_{ip}(BT_H),E_T(BT_H)]
\end{split}\label{TTH-L-BTAG}\\
\intertext{}
\begin{split}
 {\it\bf L\_KINE} &      = \left[ 1 - \left(\frac{E_T(B_1H,B_2H,BT_L,l,\nu,BT_H,J_1W,J_2W)}{E_T(B_1H,B_2H)+E_T(BT_L,l,\nu)+E_T(BT_H,J_1W,J_2W)}\right)\right]^{10}\\
                  & \times \left[\frac{\sum_{i = 1,2} E_T(B_iH) + \sum_{i = 1,2} E_T(BT_i) + E_T(l) + \sum_{i = 1,2} E_T(JW_i)}{E_T^{tot}(\text{ECAL+HCAL+VFCAL})} \right]^3\\
                  & \times \left[ \frac{E_T(B_1H,B_2H)}{E(B_1H,B_2H)} \times \frac{E_T(BT_L,l,\nu)}{E(BT_L,l,\nu)} \times \frac{E_T(BT_H,J_1W,J_2W)}{E(BT_H,J_1W,J_2W)} \right]^{0.1}
\end{split}\label{TTH-L-KINE}
\end{align}


%
\end{document}